\documentclass[conference]{IEEEtran}
\IEEEoverridecommandlockouts
\usepackage{cite}
\usepackage{amsmath,amssymb,amsfonts,amsthm}
\usepackage{algorithmic}
\usepackage{graphicx}
\usepackage{multicol, blindtext}
\usepackage{textcomp}
\usepackage{svg}
\usepackage{subfig}
\usepackage{amsmath}

\graphicspath{ {/} }
\usepackage{color} 
\usepackage{float}


\def\BibTeX{{\rm B\kern-.05em{\sc i\kern-.025em b}\kern-.08em
T\kern-.1667em\lower.7ex\hbox{E}\kern-.125emX}}

\IEEEoverridecommandlockouts

\begin{document}

\title{\fontsize{23.7}{26}\selectfont
Dynamic Standalone Drone-Mounted Small Cells
}

\author{
\IEEEauthorblockN{Igor Donevski, Jimmy Jessen Nielsen
\IEEEauthorblockA{Department of Electronic Systems, Aalborg University, Denmark}}
\IEEEauthorblockA{e-mails: igordonevski@es.aau.dk, jjn@es.aau.dk}

} 

\maketitle

\begin{abstract}

This paper investigates the feasibility of Dynamic Horizontal Opportunistic Positioning (D-HOP) use in Drone Small Cells (DSCs), with a central analysis on the impact of antenna equipment efficiency onto the optimal DSC altitude that has been chosen in favor of maximizing coverage. We extend the common urban propagation model of an isotropic antenna to account for a directional antenna, making it dependent on the antenna's ability to fit the ideal propagation pattern. This leads us to define a closed-form expression for calculating the Rate improvement of D-HOP implementations that maintain constant coverage through antenna tilting. Assuming full knowledge of the uniformly distributed active users' locations, three D-HOP techniques were tested: in the center of the Smallest Bounding Circle (SBC); the point of Maximum Aggregated Rate (MAR); and the Center-Most Point (CMP) out of the two aforementioned. Through analytic study and simulation we infer that DSC D-HOP implementations are feasible when using electrically small and tiltable antennas. Nonetheless, it is possible to achieve average per user average rate increases of up to 20-35\% in low user density scenarios, or 3-5\% in user-dense scenarios, even when using efficient antennas in a DSC that has been designed for standalone coverage.

\end{abstract}


\section{Introduction}

Drone, a.k.a., UAV (Unmanned Aerial Vehicle) usage has excelled in the last decade due to commercial demand for consumer uses such as: photography, entertainment and payload delivery, or public service uses such as search and rescue. In the world of wireless communications, because of their mobility and modularity, the flying devices are considered useful as drone mounted access points that provide or improve localized communication quality. In accord, the need for airborne base stations has been accentuated in the last five years, as it can be noticed from the overabundance in scientific and standardization activity \cite{mainsurvey,tutorial}. The concept has been identified as useful in diverse use cases: disaster recovery missions, failures of the main infrastructure, coverage assistance \cite{dynamic} for traffic surges, or sinks for Internet of Things (IoT) devices \cite{iot}. 

\begin{figure}[t]
\centering
\includegraphics[width=1\columnwidth]{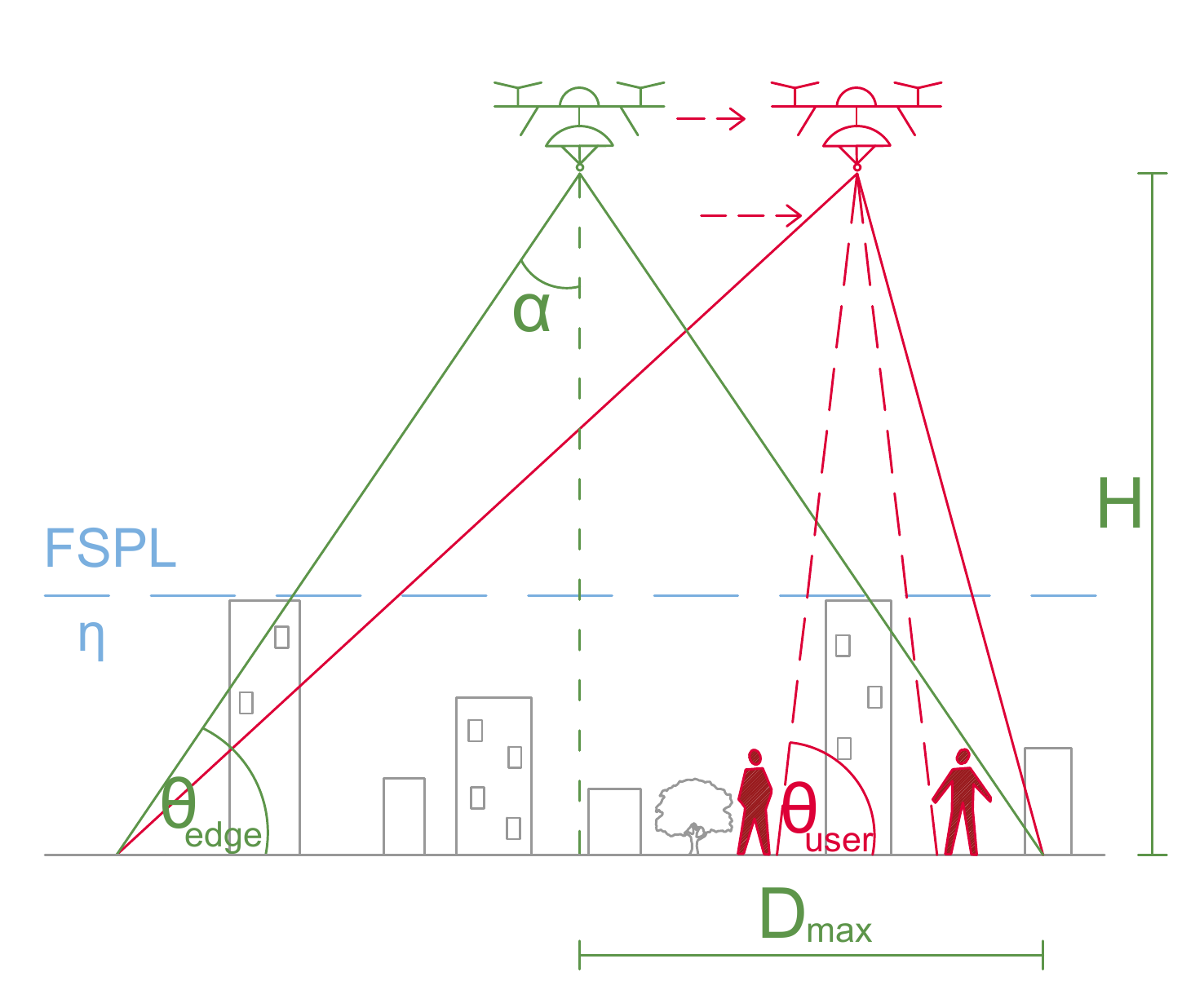}
\caption{Reference cell placement (green), and DSC with D-HOP implementation active (red).}
\label{fig:model}
\end{figure}

\subsection{Goals and Motivation}

Drones are eligible implementations of Small Cells (DSCs) that offer good coverage in urban areas. Compared to traditional cellular networks, DSCs avoid strong signal shadowing because they are positioned high relative to the user equipment. This benefit is inherent to all Low or High Altitude Platforms (LAPs and HAPs), where DSCs can be considered as a subcategory of LAPs. In accord, there has been a plethora of publications explaining and exploiting the channel improvements due to the relatively high altitude \cite{ optlap, atg, otherantenna,3dplace}.

The motivation behind using DSCs is that they offer an alternative that avoids infrastructure costs. In fact, 52\% of the Mobile Network Operator's (MNO) CAPEX is spent on site acquisition and construction; in addition to which, site rental dominates the MNO's OPEX, and is increasingly more expensive, with up to 42\% in developed countries \cite{mainsurvey}. Unfortunately, real world deployment feasibility is challenged by the weight and energy limitations to UAV air-time. Dedicated models, designed solely for this purpose should manage to minimize the impact of these drawbacks and rationalize the financial input of investing. %In that sense, we think that it is imperative to understand the type of wireless access that DSCs are inherently efficient in, and reinforce that idea such that technically adequate solutions can be developed.

For this purpose, we design a DSC system model that uses a directional antenna in combination with Dynamic Horizontal Opportunistic Positioning (D-HOP) techniques. We consider both elements of the model essential to the concept of DSCs, and not accounting for either will result in misplaced resources and/or unused opportunity. We illustrate a toy example of our approach on Fig. \ref{fig:model}. With green we show a reference deployment of a DSC with no D-HOP applied. Here, the height $H$ and radius $D_\text{max}$ are defined during the planning phase in favour of standalone coverage maximization, and are dependent on the region topography and the antenna efficiency to fit the beamwidth of $2 \cdot \alpha$. With red we show how the DSC would apply D-HOP in attempts of avoiding shadowing for two active customers, all the while, it tilts the antenna to maintain a constant coverage of the initially designated area. Our model is based solely on the likelihood of encountering a strong shadowing effect for a specific type of terrestrial topology instead of having full knowledge of the regional infrastructure; since we expect deployments like this need to be easy to deploy, and apply to natural disaster scenarios as well\cite{DMs}. %\PP{OK, but this example assumes that the drone knows exactly the positions of the users as well as which user is shadowed in each drone position, such that the drone can choose a position where there is no shadowing. These are strong assumptions and I do not see them used further in the  paper.}

\subsection{Relation to State of the Art}

This is the first work to investigate the combination and compatibility of: using a directional transmitter, the impact of the antenna's efficiency in fitting the beamwidth requirements, and most importantly, its impact on D-HOP improvements for active users. Within this work, we test the performance of three D-HOP techniques, and discuss on choosing the most adequate one for Standalone DSCs. As such, the antenna efficiency has a central role to D-HOP usefulness. The works of \cite{DCs,JointDCs,CellConn}, although considering directional antennas, do not account for the impact of the efficiency of the transmitter, are mainly concerned with 3D placement of the drone and omit analysis on the benefits of different D-HOPs. On the other hand, the works in \cite{dynamic} and \cite{3dplace} are mainly concerned with the location of the drone, assume that they do operate in a non-standalone manner, and omit the impact of directional transmitters altogether. With this, we hope to introduce the reader to the potential benefits and implementation complexities that concern deployment of D-HOP enabled standalone DSCs. 

The paper is organized as follows.
Sec. \ref{model} describes the considered RAN scenario, explains the directional antenna model, and produces a closed-form expression of the D-HOP model. 
Sec. \ref{dynamicity} explains the tested D-HOP techniques.
Sec. \ref{numerical} presents numerical results from the performed simulations, and discusses the outcomes. 
Finally, Sec. \ref{Conclusions} concludes with the impact of this work.

\section{System Model}
\label{model}

%The proposed model, as sketched on Fig. \ref{fig:model}, was decided upon several assumptions. 
We begin by assuming that the backhaul link is over-provisioned, and does not have any additional requirements that need to be accounted for. To evaluate the effectiveness of the positioning algorithms in a scenario where the DSCs act as solitary wireless service nodes, we need to assume that the equipment successfully maintains coverage at all times over the whole designated cell with radius $D_\text{max}$ by mechanically or electronically tilting the antenna.
%With this said, we continue in analyzing the influence that the scenario's geometry and the dynamic repositioning techniques have onto the expected signal strength at each active user. 

\subsection{Propagation Model}

Looking at Fig. \ref{fig:model} we notice that buildings may obstruct a user's direct link towards the DSC. We therefore consider the user as belonging to one of two propagation groups, users with Line of Sight (LoS) and No LoS (NLoS) \cite{optlap}. 

%\PP{You need to make the model precise. For a }

To express the likelihood of a user device belonging to either of the propagation groups we require a model for the LoS probability. In this service, the ITU has created a model \cite{ITU} that can be approximated to an s-curve defined by two topology constants $a$ and $b$ as a function of the elevation angle at user side $\theta_\text{user}$ that is expressed in degrees $ 0 \leq \theta_\text{user} \leq 90$. The model defined in Eq. \eqref{eq:plos}, provides good approximations of the ITU model for calculating $ P(\text{LoS})$, that is especially accurate for urban scenarios \cite{atg}.

\begin{equation}
\label{eq:plos}
    P(\text{LoS})  =\frac{1}{1+a\exp(-b(\theta_\text{user}-a))}
\end{equation}

%Having defined the propagation groups and the probability of belonging to them, we need to define the expectation of the path loss for each. 
As shown on Fig. \ref{fig:model} we define the total path loss as a combination of Free Space Path Loss (FSPL) and the expected shadowing coefficient for each of the propagation groups $\eta_\text{LoS}$ and $\eta_\text{NLoS}$. These values represent the means of the normally distributed excessive path loss, that is induced due to the large features of the topology in LoS and NLoS, respectively. Assuming a directional antenna with directivity measure $D_{t}$ is mounted on the drone, we can define the path loss per propagation group as: %
\begin{equation}
\label{eq:pllos}
 L_\text{LoS}  = - 10\log{(D_{t})} + 20\log{(d)} + 20\log{(\frac{f4\pi}{c})} + \eta_\text{LoS} 
\end{equation}
\begin{equation}
\label{eq:plnlos}
    L_\text{NLoS}  = - 10\log{(D_{t})} + 20\log{(d)} + 20\log{(\frac{f4\pi}{c})} + \eta_\text{NLoS}
\end{equation}

Knowing $P(\text{LoS})$ and $P(\text{NLoS}) = 1 - P(\text{LoS})$, we can continue to find the expected path loss $\Lambda$ as:
\begin{align}
\label{eq:expected}
    10\log(\Lambda)  &= L_\text{LoS} \cdot P(\text{LoS}) + L_\text{NLoS} \cdot P(\text{NLoS}) \\
      &= P(\text{LoS}) (\eta_\text{LoS}-\eta_\text{NLoS}) + L_\text{NLoS}
\end{align}

\subsection{Antenna Gain}

Going back to Fig. \ref{fig:model} we observe that the usage of a directional antenna at drone-side requires that we fix the proportions of our main lobe to fit the size of the apex angle $2\cdot\alpha$, which fixes the value of $\theta_\text{edge}$ as well. This is done with two main arguments. Firstly, a transmitter limited to its own cell will not contribute to the interference in other cells, therefore diminishing its negative impact. Secondly, using a directional antenna is a practical way of boosting the signal strength within the designated area, while ignoring the users outside the defined borders. In this way, a combination of multiple DSCs can be used, while avoiding strong inter-cell interference, as shown in \cite{otherantenna}.

We first analyze the directivity $ D_{t}$ element in eqs. \eqref{eq:pllos} and \eqref{eq:plnlos} as an ideal antenna $ D_{I}$  that perfectly covers the designated circular area. We define a value $ 0 \leq E_r \leq 1 $ that is dependent on the type, manufacturing and quality of the antenna, to measure the efficiency of our implemented antenna in reference to the ideal one $D_t = D_{I}^{E_r}$ for our purpose. This measures the strength of the Main Lobe with relation to the spread outside the assigned coverage area due to sidelobes, inadequate main lobe size or other imperfections. The most adequate antenna type for our application includes, but is not limited to, phased array antennas, as they are able to quickly adjust the direction of the main lobe. From here we first calculate $D_{I}$ as \cite{antennaest}: 
\begin{equation}
\label{eq:idealant}
    D_{I}  = \frac{4\pi}{\Omega}
\end{equation}

$ D_{I}$ is defined by a beamwidth defined by the solid angle of a perfect cone as $\Omega = 2\pi(1-\cos{(\alpha)})$. The angle $\alpha$ is half of the apex angle in either of the two two-dimensional propagation planes, as shown in Fig. \ref{fig:model}. By further applying simple trigonometry to our UAV scenario, we reach Eq. \eqref{eq:dt} as the final metric, which is dependent on the elevation angle on the edge of the cell $ \theta_\text{edge}$.
\begin{equation}
\label{eq:dt}
    D_{I}  = \frac{2}{1-\sin{(\theta_\text{edge} \frac{\pi}{180} )}}
\end{equation}

Including this in the final equation for expectation of path loss and expressing all in terms of $D_\text{max}  > 0$, $\theta_\text{user}$ and $\theta_\text{edge}$ expressed in degrees we get:
\begin{multline}
\label{eq:PLmax}
 10\log(\Lambda) =\\ \frac{\eta_\text{LoS}-\eta_\text{NLoS}}{1+a\exp(-b[\theta_\text{user}-a])} + 20\log(\frac{D_\text{max}}{\cos{(\theta_\text{user}\frac{\pi}{180})}}) \\  - {E_r}10\log{(\frac{2}{1-\sin{(\theta_\text{edge}\frac{\pi}{180})}})} + 20\log{(\frac{f4\pi}{c})} + \eta_\text{NLoS}
\end{multline}

\subsection{Cell Size and $ \theta_\text{edge}$}
We define the cell size by assuming a maximally allowed expected path loss at the point of worst case coverage when the user is located at the cell edge $\theta_\text{user} =\theta_\text{edge} = \theta$. We analyze the communication in an information theoretic manner and define the expected rate as in Eq. \eqref{eq:avgrate} in terms of bits/symbol. Since we are maximizing the cell coverage we are interested in meeting the average rate requirement: 
\begin{equation}
\label{eq:avgrate}
R_\text{avg} = \log_2(1+\frac{P_\text{tx}}{N_0\Lambda})
\end{equation}

In the following we assume for simplcity that $R_\text{avg}=1$, which gives that the drone needs to be positioned at the point where $\Lambda = \frac{P_\text{tx}}{N_0} $, and we maximize the radius of the cell $D_\text{max}$ in favor of coverage. In this way, the elevation angle at the edge of the cell is chosen to constrain the radiation radius to the planned radius of the cell but also makes sure to avoid discriminating the users located at the edge of the cell. Going back to Eq. \eqref{eq:PLmax} to find the maximal radius as a function of the elevation angle $\frac{dD_\text{max}}{d\theta}$, we conclude that it is only dependent on the scenario topography relative to our operating frequency and the efficiency of our antenna, as shown in the derivation in Eq. \eqref{eq:dtderiv}.
\begin{multline}
\label{eq:dtderiv}
    0 = \frac{\pi \tan{(\theta\frac{\pi}{180}})}{9\log(10)} + \frac{a\,b A \exp(-b(\theta -a)) }{a\exp(-b(\theta-a)+1)^2 }
    \\- {E_r} \frac{\pi  \cos{(\theta\frac{\pi}{180})}}{18\log(10)(1-\sin{(\theta\frac{\pi}{180})} )}
\end{multline}

\begin{figure}[h]
\centering
\includegraphics[width=1\columnwidth]{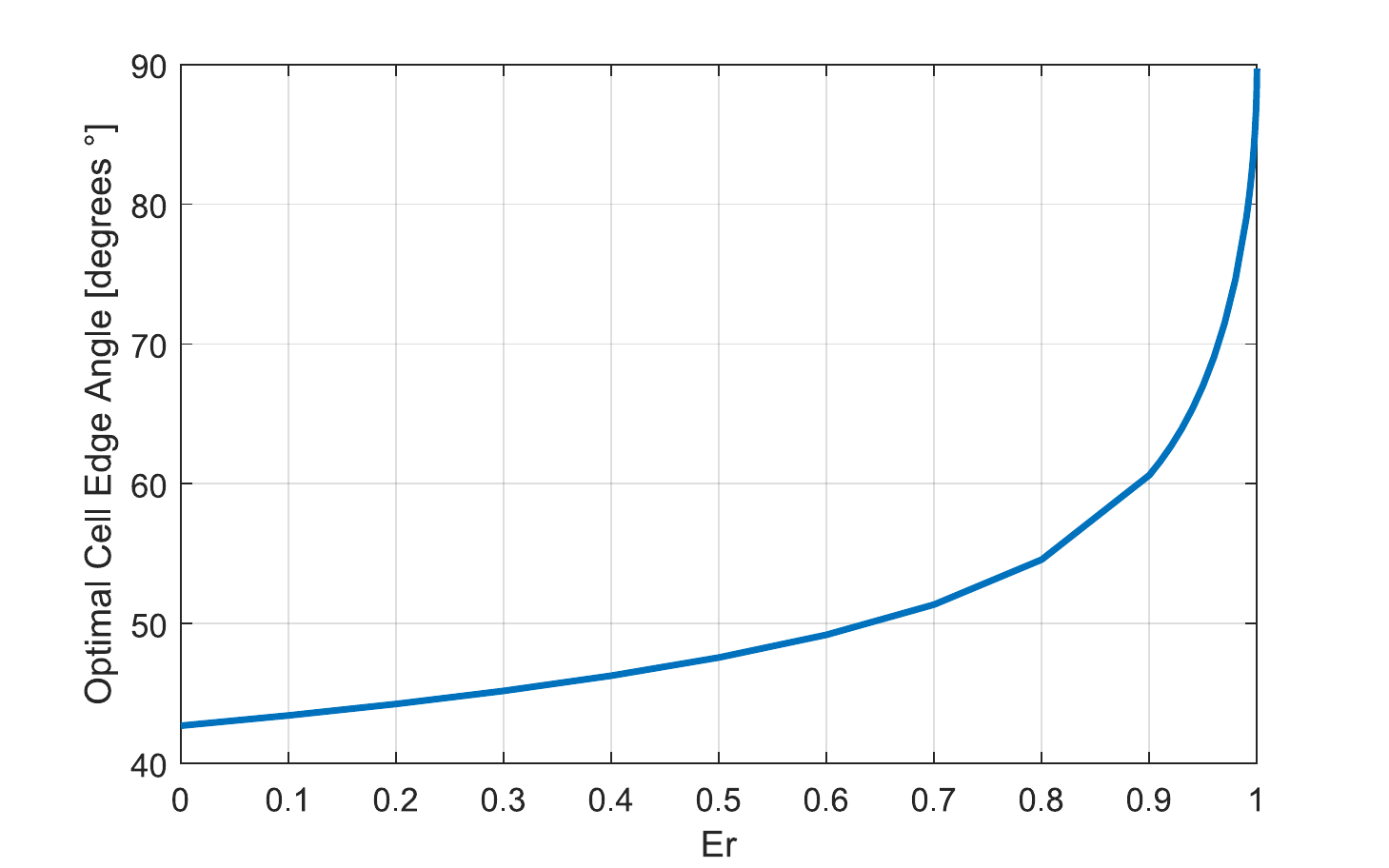}
\caption{The optimal cell edge elevation angle $\theta_\text{edge}$ as a function of $E_r$ for Urban Scenario parameters given in Sec. \ref{numerical}.}
\label{fig:evolution}
\end{figure}

Analysing eq. \eqref{eq:dtderiv} from the perspective of an urban scenario we investigate the impact of $E_r$ onto the optimal $\theta$. On Fig. \ref{fig:evolution} we follow the evolution of $\theta_\text{edge}$ in the cases of different antenna radiation efficiency. The two extremes set for $E_r$ are 0 for an ideal isotropic antenna and 1 for an ideal conical antenna. Having  $E_r=0$ coefficient cancels the $D_t$ member in Eq. \eqref{eq:dtderiv}  and the final $\theta_\text{edge}$ is identical to the one of using an isotropic antenna. Having $E_r=1$ breaks the point of optimality in Eq. \eqref{eq:dtderiv}. In other words, when possessing an ideal transmitter we should be able to establish point to point transmissions with infinitely big gain, meaning that $E_r$ should be strictly smaller than 1 \cite{antennaest}. This dependence of $\theta_\text{edge}$ on $E_r$ defines the principal contribution of our model, and it shows the important details that MNOs need to pay attention to when designing the geometry of the system.

\subsection{User Data Rate}
In a circular coverage area with radius $D_\text{max}$ there are $N$ active users, each i$^{\text{th}}$ user's location within the area is defined with two coordinates $(x_i,y_i)$, where $i \in {1,2 ... N}$, and the drone occupies a position with coordinates $(x_\text{d},y_\text{d})$. We define each user's horizontal distance to the drone with $d_i = \sqrt{(x_i - x_\text{d})^2 + (y_i - y_\text{d})^2}$ and define a scalar $\kappa_i$ that represents the distance in relation to the cell radius $ d_i = \kappa_i \cdot D_\text{max}$, where $2 \geq \kappa_i \geq 0$. We reformulate our expected path loss in eq. \eqref{eq:PLmax} in terms of $\kappa_i$, as:
\begin{multline}
\label{eq:distr}
 10\log(\Lambda)  = \frac{\eta_\text{LoS}-\eta_\text{NLoS}}{1+a\exp(-b[\theta_\text{user}-a])} \\ + 20\log(\sqrt{\kappa_i^2+\tan(\theta_\text{edge}\frac{\pi}{180})^2}) + 20\log(D_\text{max}) + C
\end{multline}

Where $\theta_\text{user}$ is dependent on $\kappa_i$ and is $\theta_\text{user}=\arctan(\frac{h}{\kappa_i D_\text{max}})=\arctan(\frac{\tan(\theta_\text{edge}\frac{\pi}{180})}{\kappa_i})$ ; and parameters that are independent of $\kappa_i$ or $D_\text{max}$ are $C = 20\log{(\frac{f4\pi}{c})} + \eta_\text{NLoS} - E_r10\log{(D_{I})} $. Additionally, eq. \eqref{eq:distr} is continuous over the whole range of possible user positions as well as $\kappa_i\geq0$. For convenience, we group all members of the equation that depend on $\kappa_i$ to define the horizontal repositioning gain $G_\text{pos}$ as:
\begin{align}
\label{eq:gk}
G_\text{pos}(\kappa_i) &= \frac{\eta_\text{LoS}-\eta_\text{NLoS}}{1+a\exp(-b[\theta_\text{user}-a])}\\  &+ 10\log(\kappa_i^2+\tan(\theta_\text{edge}\frac{\pi}{180})^2)
\end{align}

And receive a final, closed form equation for the expected pathloss of user at distance $\kappa_i$ in a cell with radius $D_\text{max}$ as:
\begin{multline}
\label{eq:final}
 10\log(\Lambda(\kappa_i,D_\text{max}))  =  G(\kappa_i) + 20\log(D_\text{max})\\ + 20\log{(\frac{f4\pi}{c})} + \eta_\text{NLoS} - E_r10\log{(\frac{2}{1-\sin{(\theta_\text{edge} \frac{\pi}{180} )}})}
\end{multline}

This results in the per-user expected rate being:
\begin{align}
\label{eq:rate}
R_i &= \log_2(1+{\frac{\Lambda(\kappa_i=1,D_\text{max})}{\Lambda(\kappa_i,D_\text{max})}})  \\ 
&= \log_2(1+10^\frac{G_\text{pos}(\kappa_i=1)-G_\text{pos}(\kappa_i)}{10})
\end{align}

\begin{figure}[t!]
\centering
\includegraphics[width=0.8\columnwidth]{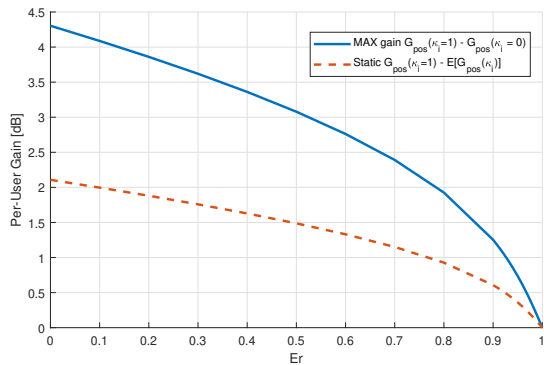}
\caption{Analysis on maximum possible D-HOP gain as a function of $E_r$.}
\label{fig:gaineffdristr}
\end{figure}

From the involved parameters, it is obvious that the benefits from D-HOP implementation depend on the cell's geometrical proportions and not on its absolute size. Additionally, on Fig. \ref{fig:gaineffdristr} we show how the span of possible D-HOP gains evolve as a consequence of the behaviour shown in Fig. \ref{fig:evolution} that is dependent on the antenna's efficiency. Since all aforementioned parameters are predefined when planning the communications model, in the next section we focus on lowering the values for $\kappa_i$ through the means of D-HOP.

\section{D-HOP of Drone Small Cells}
\label{dynamicity}

We open this section by stressing that Drone D-HOP mobility does not affect the coverage area; as it can adjust its antenna angle (physically/electronically) to fully cover the area with radius $D_\text{max}$, and inactive users in the area can announce their location and activation time with a rate $R_\text{acc} \ll R_\text{avg}$. Distortions of the circular coverage field to an oval one due to the angle of the transmitter are considered to be negligible. From here, our goal is to improve the channel condition for the active users in the area without neglecting new requests. We do this by performing constant dynamic movements in the horizontal plane with height $H = D_\text{max}\tan(\theta_\text{edge})$.

\begin{figure}[t!]
\centering
\includegraphics[width=1\columnwidth]{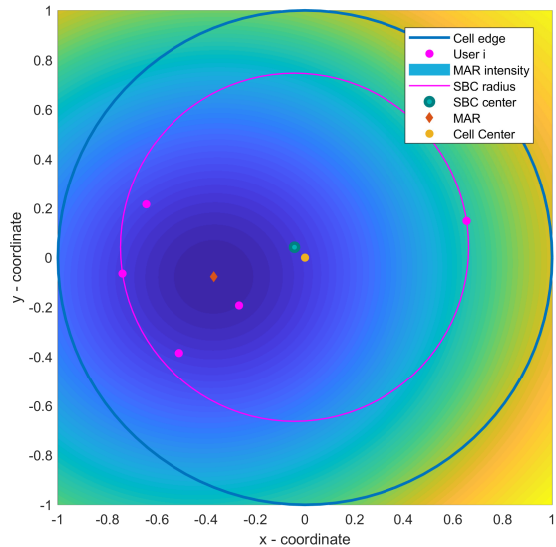}
\caption{An illustration of a single D-HOP scenario, over MAR intensity map.}
\label{fig:scenario}
\end{figure}

In an arrangement of user locations our goal is to optimize the drone position $(x_\text{d},y_\text{d})$ and we identify two significant points, one performance maximization oriented, and the other as fairness oriented. We illustrate this in Fig. \ref{fig:scenario}.

%From here, calculating the desired horizontal placement, in a scenario where all end users’ locations is straightforward. We have to note that in a real-world application, the locations would need to be anticipated with some uncertainty, which is expected to add some computational complexity strain but since the nature of this analysis is stohastic, slight miscalculations should not create dramatic . 

\subsection{Smallest Bounding Circle (SBC)}
One approach that carries great geometric significance, is to set the DSC location in the center of the minimum bounding circle of all active users, which is the smallest circle that contains all points inside \cite{3dplace}. With this, the goal is to maximize the fairness of the dynamic system by minimizing $d_\text{f}  = \max[d_i]$ $\forall$ $i$. The minimum bounding circle is a well known computational geometry problem falling under the umbrella of facility location, or the 1-center problem. 

\subsection{Maximum Aggregated Rate (MAR)}
The second position of geometric significance is where drone placement would achieve minimal total distance to all active users Min: $d_\text{m}  = \sum_{i} {d_i}$, or, the centroid of all points. Although the SBC $d_\text{f}$ is a universal fairness maximization approach, the centroid $ d_\text{m}$ is not, and we substitute it for a more adequate performance parameter. In its stead we use the aggregate rate improvement, as: 
\begin{equation}
\label{eq:MAR}
\max_{x_\text{d},y_\text{d}} \sum_{i=0}^{N}R_i 
\end{equation}

\subsection{Center-Most Point (CMP)}
Finally, limitations regarding the mobility of the UAV need to be taken into account as it cannot instantaneously relocate on every position with every shift in user behaviour. This requires inspecting a more travel distance conservative repositioning technique. In favor of this, we create a repositioning algorithm that puts the drone either at maximum gain, or maximum fairness, depending on which of both points is the Centermost Point. This is done knowing that if averaged over an infinite amount of users and timeslots, the optimal position of the drone is in the very center.

\begin{figure*}[t!]
\centering
\includegraphics[width=1\linewidth]{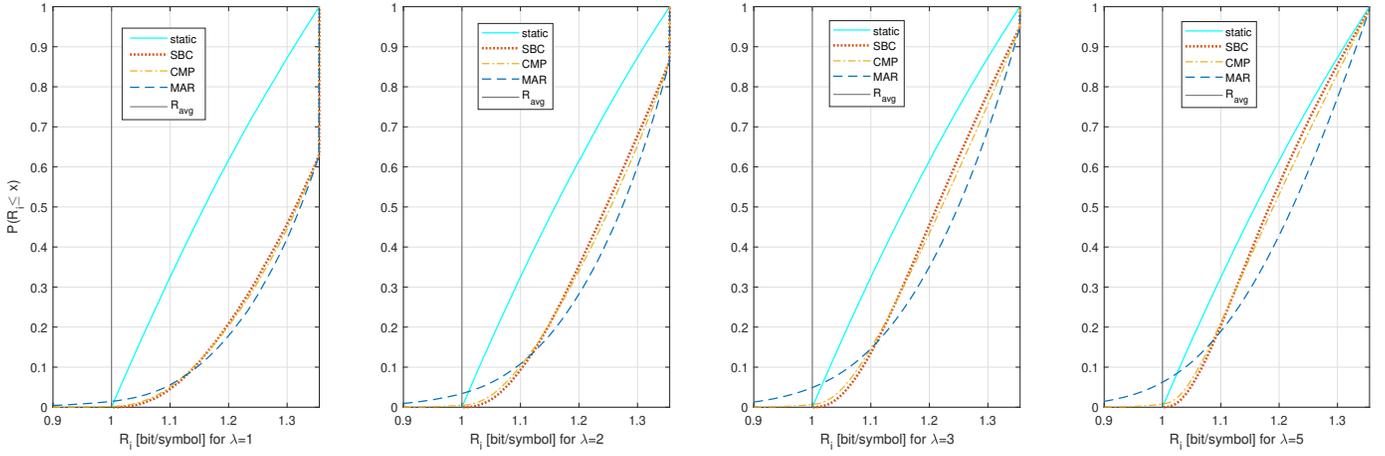}
\caption{The CDF of the probability for a user to receive rates higher than $R_\text{avg}$ tested for four different user densities.}
\label{fig:comparegain}
\end{figure*}

\section{Simulation and Results}
\label{numerical}

We consider a snapshot based, simplistic and replicable testing scenario where the positioning of the UE occurs in timeslots, and each timeslot has no correlation to the previous one. No assumptions are done with user mobility in mind, and each user can be uniformly located within the cell's limits. We test the system under the Urban scenario conditions, with parameters: $a=9.61, b=0.16 ,\eta_\text{LoS}=1,\eta_\text{NLoS}=20, f=2~\text{GHz} $\cite{optlap}. 

On Fig. \ref{fig:gaineffdristr} we show that $E_r$ has a strong impact on the feasibility of D-HOP implementations. Moreover, dynamic DSCs make more sense in use cases that require antennas with lower $E_r$. This is due to the dependency of the cell's edge optimal elevation angle $\theta_\text{edge}$ to the $E_r$ coefficient. Luckily, this goes in favor of dynamic DSCs since lighter, cheaper, and electrically small antennas are expected at drone side. We, therefore arbitrarily set $E_r=0.6$ as an example of an adequately chosen antenna guided by the comments of \cite{antennaest}.
 
\begin{figure}[b]
\centering
\includegraphics[width=1\columnwidth]{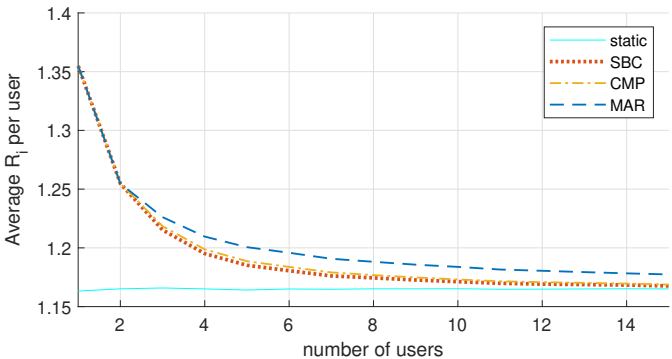}
\caption{Per user expected rate, as a function of the number of active uniformly distributed users.}
\label{fig:usernumberavg}
\end{figure}

We first accentuate the importance of the expected geometric location of the users and that D-HOP finds purpose in unbalanced user formations, especially in the existence of clusters \cite{dynamic}.
However, we do not account for user clustering, and we assume that if averaged over an infinite amount of timeslots, the optimal location for positioning the drone is in the very center of the cell such as in a well planned cell placement. As a consequence, in the case when the drone takes no action and stays in the center of the cell, we still detect average rate improvements, with regards to the preset average rate $R_\text{avg}$, since the users are not always located at the cell's edge. This gain, shown on Figs. \ref{fig:usernumberavg} and \ref{fig:gaineffdristr}, is entirely due to the user location distribution, and is imperative for evaluating the usefulness of any repositioning technique.

From the simulation results at Fig. \ref{fig:usernumberavg} it is obvious that the more evenly distributed active users are present, the less we exploit the D-HOP advantages. Special cases consist of only two or less active users, where any of the proposed solutions would behave the same. Therefore, Dynamic DSC deployments are well suited for areas with low user density occurrences, and in the best cases can increase the expected per user rates by 17\% with regards to static, or 34\% for users at the cell edge. Which is a good result considering that we avoid changing the drone's altitude, in service of avoiding coverage holes.

In Fig. \ref{fig:comparegain}, we show how each D-HOP approach modifies the distribution of the distances between the users and the drone and therefore achieve higher rates for most users. Here we imitate four instances of Poisson Point Process (PPP) where every timeslot has $N$ users that are Poisson distributed $N\sim\text{Poisson}(\lambda)$. It is immediately noticeable that the MAR approach has the best performance on average, at the expense of putting roughly 5\% of its users at distances $\kappa_i>1$. Nonetheless, it may be feasible for systems with no hard coverage constraints since when compared to the static it improves the rates by 5.6\% for the average user and offers 21.5\% increase over the preset average rate. The CMP technique diminishes the impact of users at distances bigger than the Radius of the cell, however, it still does not evade all $\kappa_i>1$ situations. The SBC approach obviously avoids having such cases at all, while it also improves the expected rate for the lowest fifth percentile of our users by 3\% for $\lambda=5$, and up to 10\% for $\lambda=1$. This makes the bounding circle ideal for DSCs that offer high reliability for all users within the cell.
 
\begin{figure}[t]
\centering
\includegraphics[width=1\columnwidth]{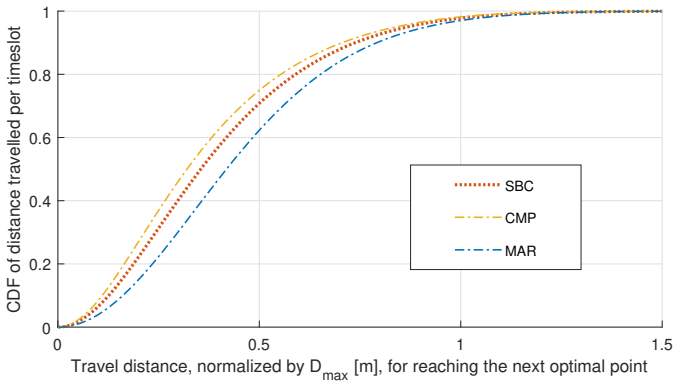}
\caption{Per timeslot drone travel distance CDF, normalized by $D_\text{max} $ in $ [m]$, for user density of $\lambda = 5$.}
\label{fig:comparedistance}
\end{figure}

To account for the limitations of the equipment, we investigate the distributions for displacement requirements per D-HOP. In Fig. \ref{fig:comparedistance} we show that the distances travelled for the SBC and the CMP have an obvious advantage over the MAR approach. The advantage of using the CMP technique with regards to mobility requirements makes it an adequate solution for less rigorous reliability requirements, as it offers a balance between fairness and rate maximization mobility requirements for the drone. 

Deriving from Figs. \ref{fig:gaineffdristr} and \ref{fig:comparegain}, and eq. \eqref{eq:final} we conclude that the span of possible performance gains are predetermined by the system topology and $E_r$, while the user distribution and D-HOP technique impact the gains achieved within that span. Since the expected D-HOP gains do not scale with the cell size $D_\text{max}$, its implementation is expected to have a higher impact in smaller cells. Additionally, drone travel distance requirements will demand much lesser drone speeds for smaller $D_\text{max}$.

Both points reinforce the fact that DSC repositioning benefits from the added complexity in the cases of small or pico cells. For the existence of reliable communications we can conclude that the predefined parameters regarding the radius/altitude of the UAV position or the angle of coverage, are very strict, and fulfill the stringent link budget constraints to establish the desired reliability \cite{all, urllc}. Therefore, following this model, we ascertain the D-HOP DSC's inherent eligibility in reliability applications by calling for user-fairness techniques as most effective D-HOP implementations for Drone Small Cells.

\section{Conclusions} 
\label{Conclusions}
This work investigates the concept of DSCs and accounts for exploiting all its advantages. We note the importance of knowing the efficiency of the available antenna equipment as it directly influences the optimal geometry of the model. We then quantify its impact over the dynamic repositioning gains, and conclude that the gains achieved from repositioning are mainly beneficial when using small antennas. For the tested urban scenario, we achieve per user average rate improvements of up to 20-35\% in low-user density scenarios, or 3\% - 5\% in dense scenarios. Which for our model is extremely well considering we assume balanced and uniform user positions with standalone and constant coverage over the whole area.
% Therefore, it is our final verdict that the added complexity of making the DSCs dynamic should be done in service of reliability; which in turn, justifies picking a fairness/reliability maximizing dynamic repositioning technique, the bounding circle, as the most adequate one.

\section*{Acknowledgment}
The work was supported by the European Union's research and innovation programme under the Marie Sklodowska-Curie grant agreement No. 765224 ''PAINLESS'' within the Horizon 2020 Program.

The authors would like to thank prof. Petar Popovski for his valuable guidance.

\end{document}